\documentclass[12]{article}
\usepackage[utf8]{inputenc}
\usepackage{authblk}
\usepackage[nottoc]{tocbibind}
\usepackage{graphicx}
\usepackage{commath}
\usepackage{bm}
\usepackage{amsmath}

\DeclareMathOperator*{\argmin}{arg\,min}
\usepackage{caption}
\usepackage{array}
\usepackage{hyperref}
\usepackage[margin=1.25in]{geometry}

\title{A statistical approach for controlling the probability of false alarm and missed detection in smartphone-based earthquake early warning systems}

\author[1]{Frank Yannick Massoda Tchoussi}
\author[2]{Francesco Finazzi}

\affil{Department of Economics, University of Bergamo, via dei Caniana, 2, Bergamo, 24127, Italy}

\begin{document}
\maketitle

\begin{abstract}
Smartphone-based earthquake early warning systems (EEWS) are emerging as a complementary solution to classic EEWS based on expensive scientific-grade instruments.
Smartphone-based systems, however, are characterized by a highly dynamic network geometry and by noisy measurements. Thus the need to control the probability of false alarm and the probability of missed detection. 

This paper proposes a statistical approach based on the maximum likelihood method to address this challenge and to jointly estimate in near real-time earthquake parameters like epicentre and depth.

The approach is tested using data coming from the Earthquake Network citizen science initiative which implements a global smartphone-based EEWS.
\end{abstract}


\section{Introduction}
Earthquake early warning systems (EEWSs) \cite{satriano2011earthquake,cremen2020earthquake} are deployed in seismic areas to detect earthquakes in real-time, in order to send a forewarning to citizens and to stop critical processes before the ground shaking begins.
Classic EEWSs are based on a dense network of scientific-grade instruments with construction and operational costs in the order of millions of euros \cite{given2014technical}.

As a result of the accessibility of smartphones, and the advancement of technologies for mobile apps development, low-cost EEWSs can be implemented. This path has been explored by the Earthquake Network (EQN) citizen science initiative \cite{finazzi2016earthquake, finazzi2020earthquake, bossu2022shaking}, which, since 2013, implements the first smartphone-based EEWS.

Smartphones are used to detect the ground shaking induced by the earthquake and a warning is issued as soon as the earthquake is detected. People living at a further distance from the epicentre may be alerted before they are reached by the damaging seismic waves. 

The primary challenge faced by the EQN is to control the probability of false alarm and the probability to miss an earthquake. Since the earthquake detection is based on smartphone accelerometer readings, alerts may be triggered by events unrelated with earthquakes. Also, it is possible that the network misses a (possibly strong) earthquake, especially if the number of monitoring smartphones is small. Both false alarms and missed detections undermine people trustiness on the EEWS.

This paper proposes a statistical approach based on maximum likelihood and hypothesis testing methods to address the above mentioned problems and to jointly estimated earthquake epicentre and depth in near real-time (i.e., in less than one or two seconds). 

The statistical approach is developed and tested using Monte Carlo simulations which rely on the EQN smartphone network, and it is then applied to some true and false EQN earthquake detections.

\section{Detection algorithm}
Before describing the statistical methodology developed in this work, we detail the output of the earthquake detection algorithm currently implemented by EQN \cite{finazzi2017statistical}. For any given area of radius $30$ km, the algorithm compares the number of triggering smartphones in the last $10$ seconds with the number of active smartphones. A triggering smartphone is a smartphone that detected an acceleration above a threshold, while an active smartphone is a smartphone known to be monitoring for earthquakes. If the ratio between triggering smartphones and active smartphones exceeds a threshold, an earthquake is claimed to be detected. The output of the detection algorithm consist of the detection location and of the list of the triggering smartphones (triggers for short), which are identified by their spatial coordinates (latitude and longitude) and by the triggering time.

\section{Statistical modelling}\label{sim3}
The generic observed triggering time for a smartphone sensing an earthquake is modelled as
\begin{equation}
   t_{i} = t_i^* + \epsilon_{i},
\end{equation}
\newline
where $t_i^*$ is the expected triggering time while $\epsilon_{i} \sim N(0, \sigma_{\epsilon}^2)$ is a random component. More in detail
\begin{equation}\label{eq1}
    t_i^*=  \frac{D_{i,H}}{v} + t_O,
\end{equation}
with
\begin{equation}\label{dist1}
     D_{i,H} = \sqrt{d_E^2 + 4R(R - d_E)sin\left(\frac{D_{i,E}}{2R}\right)^2},
\end{equation}
the distance between the hypocentre and the smartphone location, $v$ the seismic wave speed and $t_O$ the earthquake origin time. In (\ref{dist1}), $D_{i,E}$ is the distance between the epicentre $(lat_E,lon_E)$ and the smartphone location, $d_E$ is the earthquake depth and $R$ the earth radius (6371 km).
Here, it is assumed that all smartphones either detect the primary seismic wave ($v=7.8$ km/s) or they all detect the secondary wave ($v=4.5$ km/s). This assumption is justified by the fact that the earthquake detection is based on smartphones within a radius of $30$ km, which is a relatively small area.

The role of the random component $\epsilon_{i}$ is to model the difference between the expected and the observed triggering time. This difference is mainly due to the smartphone detection delay and to a seismic wave speed that may differ from the expected values.

Let us define $\Delta t_i =  t_{i} - t_i^*$, then $ \Delta t_i \sim N(0, \sigma_{\epsilon}^2)$. The model parameter vector is $\bm{\theta} = (lat_E, lon_E, d_E, t_O, \sigma_{\epsilon}^2)$.\\
To classify an earthquake detection between true and false, we implement a statistical hypothesis test on the variance of $\Delta t_i$. The system of hypothesis is given by
 \begin{equation}\label{test}
       \begin{aligned}
           \left\{ 
            \begin{array}{c}
            H_{0}: \sigma_\epsilon^2 = \delta \\
             H_{1}: \sigma_\epsilon^2 > \delta.
            \end{array}%
            \right.
       \end{aligned}
 \end{equation}

Null hypothesis is rejected when the variance is higher than expected, namely when smartphone triggering times do not follow the propagation law of primary or of the secondary seismic wave.
The test statistic is
\begin{equation}
    T = df\frac{{\sigma_\epsilon}^2}{\delta},
\end{equation}
which, under the null hypothesis, is distributed  as a chi-square with $n - 3$ degree of freedom ($df$), where $n$ is the number of triggering smartphones and $3$ is the number of estimated parameters in equation (\ref{mle2}).\\

Since we do not know which seismic wave is detected by the smartphones, two tests are done: one with $v=7.8$ km and one with $v=4.5$ km. The detection is classified as a false earthquake if the null hypothesis is rejected under both tests. Otherwise the earthquake is classified as true.

It is worth noting that, in general, $\delta$ is unknown since the variance of $\Delta t_i$ is hard to assess under a true earthquake. In Section (\ref{sim4}), $\delta$ is estimated using Monte Carlo simulations in order to fix the desired level of type I error. Since the test is on the variance of the difference between the observed and the expected triggering time for a given value of $v$, we assume that $\delta$ does not depend on $v$.

\subsection{Model estimation}
To implement the test in (\ref{test}), an estimate of $\sigma^2_{\epsilon}$ is needed. Using maximum likelihood, elements of the parameter vector $\bm{\theta}$ are jointly estimated.

Assuming to have a list of $n$ triggering smartphones provided by the EQN detection algorithm, the log-likelihood function based on the joint probability distribution of the $\Delta t_i$, $i=1,...,n$, is
\begin{equation}\label{mle}
    l(\bm{\theta}) = -\frac{n}{2}\ln{2\pi} - \frac{n}{2}\ln{\sigma_\epsilon^2} - \frac{1}{2\sigma_\epsilon^2}\sum_{i = 1}^{n}\Delta t_i^2.
\end{equation}

Note that $\Delta t_i$ are assumed to be independent. This assumption is realistic because smartphones do not share a common clock, detection delays are independent and the detection by each smartphone is influence by local factors (e.g., where the smartphone is located, at which floor of the building, the accelerometer sensitivity and so on). 

Maximum likelihood estimates of $lat_E$, $lon_E$ and $d_E$ are given by

\begin{equation}\label{mle2}
    \argmin_{lat_E, lon_E, d_E} \sum_{i=1}^{n} \Delta t_i^2.
\end{equation}

The solution of (\ref{mle2}) cannot be obtained in closed form, due to the non-linearity of (\ref{dist1}), hence estimates are obtained via numerical optimization \cite{lange1999numerical}. As usual, to avoid local minima, the numerical optimization algorithm is run multiple times starting from random initial values for $lat_E$, $lon_E$ and $d_E$. Note that $t_O$ cannot be estimated from (\ref{mle2}) since $\Delta t_i$ do not carry information on $t_O$. Nonetheless, $t_O$ is a nuisance parameter which is not needed for classifying the EQN detection between true and false.

Finally, the maximum likelihood estimate of the variance is
\begin{equation}\label{var}
   \hat{\sigma_\epsilon}^2= \frac{1}{n}\sum_{i = 1}^{n}(\hat{\Delta t_i} - \hat{\mu})^2,
\end{equation}
where $\hat{\Delta t_i} = t_i - \hat{t_i}^*$ is computed after replacing in (\ref{eq1}) and in (\ref{dist1}) the maximum likelihood estimates, while $\hat{\mu}$ is the mean of the $\hat{\Delta t_i}$.

\section{Simulation study}\label{sim1}
Since $\delta$ in (\ref{test}) is unknown, its value is estimated using Monte Carlo simulations. The simulation is based on the EQN smartphone network in Lima (Peru). True and false earthquake detections are simulated considering the locations of $1000$ smartphones. True detections are used to estimate $\delta$ while false detections are used to assess the power of the test.

\subsection{Simulation of true detections}\label{sim2}
For simulating a true earthquake, the following aspects are taken into account: the earthquake epicentre and depth, the arrival time of the seismic wave at the smartphone locations, the earthquake detectability by the smartphone and the error on the triggering time. Lastly, we account for the fact that smartphones may detect events unrelated to the earthquake.\\

The epicentre locations ($lon_E$, $lat_E$) is simulated uniformly inside the coordinates box \sloppy$[-12.39^{\circ}, -11.74^{\circ}]$ for latitude and $[-77.17^{\circ}, -76.66^{\circ}]$ for longitude. The box encompasses the EQN network of Lima. On the other hand, the earthquake depth is simulated uniformly in the range $[0, 100]$ km independently of the earthquake epicentre.

The arrival time of the seismic wave at each smartphone location is simulated from (\ref{eq1}) assuming $t_O=0$ and $v=7.8$ km/s. Only 70\% of the smartphones are made triggering because of the earthquake. For these smartphones, the error on the triggering time is simulated from a zero mean normal distribution with variance $\sigma_\varepsilon^2=1.67$. Such variance guarantees that the 1-st and the 99-th percentiles of the error are around $-3$ and $3$ s, respectively, which are realistic values for an error on the triggering time.  

Of the remaining 30\% of smartphones which do not trigger, 6\% are made triggering at random with a triggering time uniformly generated in the range $[0, 12]$ s. This implies that, when the earthquake is detected by the EQN detection algorithm, the list of triggering smartphones may include triggers unrelated with the earthquake.

Once the list of triggering smartphones is defined and sorted by triggering time, the EQN detection algorithm is applied to the list. The algorithm stops when the detection condition is satisfied, and the sub-list of triggers that concurred to the earthquake detection is given as output.

Figure 1 shows an example of a simulated true earthquake. Two separated regions can be visually identified, one with triggering smartphones (those that concurred to the detection), and another with non-triggering smartphone not yet reached by the seismic waves.

\begin{figure}[ht!]
    \centering
    \includegraphics[width=1\textwidth]{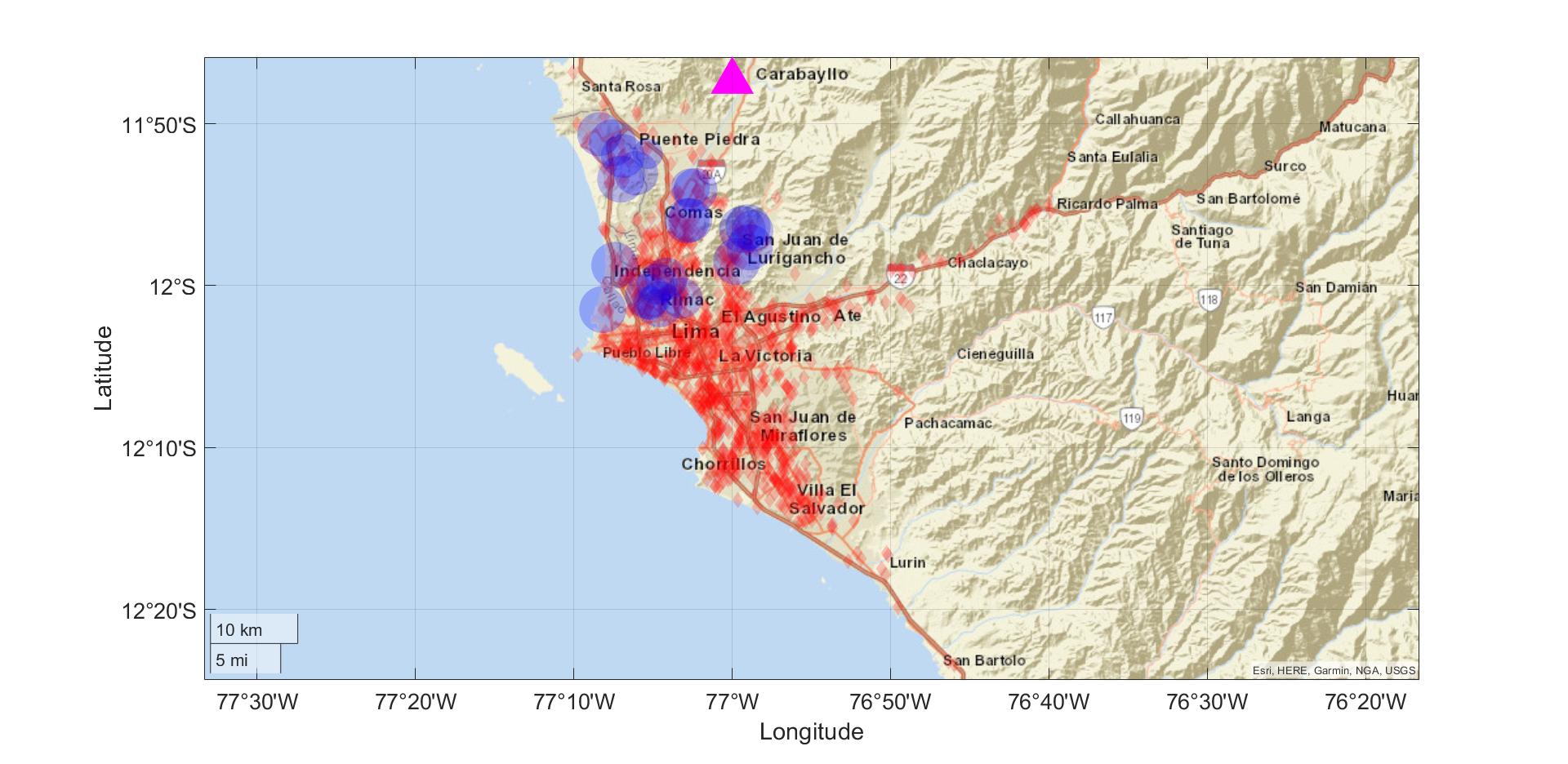}
    \captionsetup{}
    \caption{True simulated earthquake based on the EQN smartphone network of Lima (Peru). Smartphones that concurred to the earthquake detection are represented by circles with diameter proportional to the triggering time. Active smartphones are represented by diamond markers. The magenta triangular marker represents the simulated epicentre location.}
    \label{fig1}
\end{figure}

\subsection{Simulation of false detections}
To simulate a false detections, we assume that smartphones trigger at random with a triggering time which does not follow the law of the seismic wave propagation. Only 30\% of the smartphones are made triggering and their the triggering time is uniformly sampled in the range $[0, 12]$ s.

\begin{figure}[ht!]
    \centering
    \captionsetup{}
    \includegraphics[width=1\textwidth]{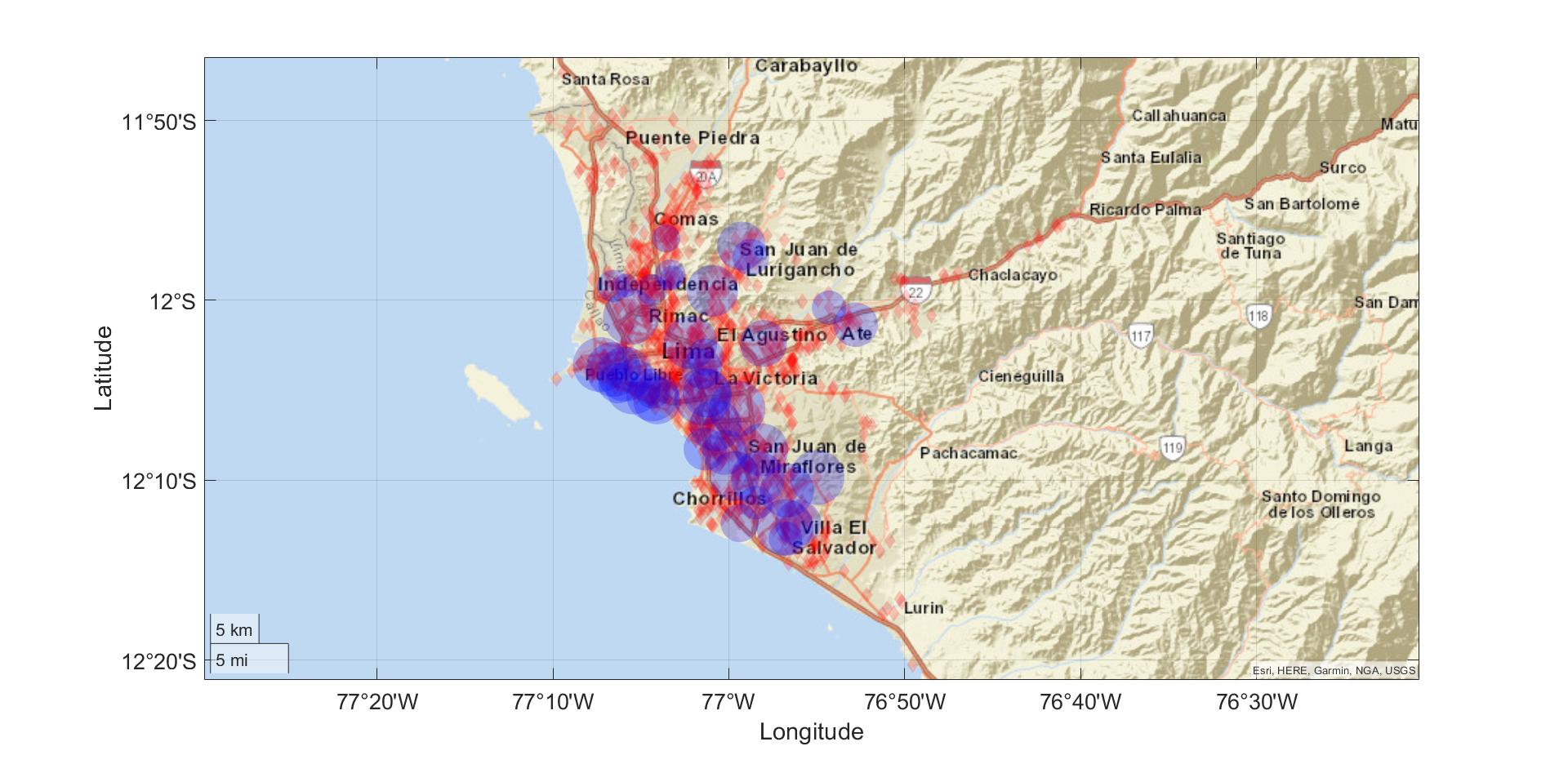}
    \caption{Simulated false detection based on the EQN smartphone network of Lima (Peru). Smartphones that concurred to the detection are represented by circles with diameter proportional to the triggering time. Active smartphones are represented by diamond markers.}
    \label{fig2}
\end{figure}

Figure \ref{fig2} shows an example of a simulated false EQN detection. Contrary to true earthquakes, no specific spatial pattern on the triggers is observed.

\subsection{Simulation results}\label{sim4}

\begin{figure}[ht!]
    \centering
     \captionsetup{}
    \includegraphics[width=1\textwidth]{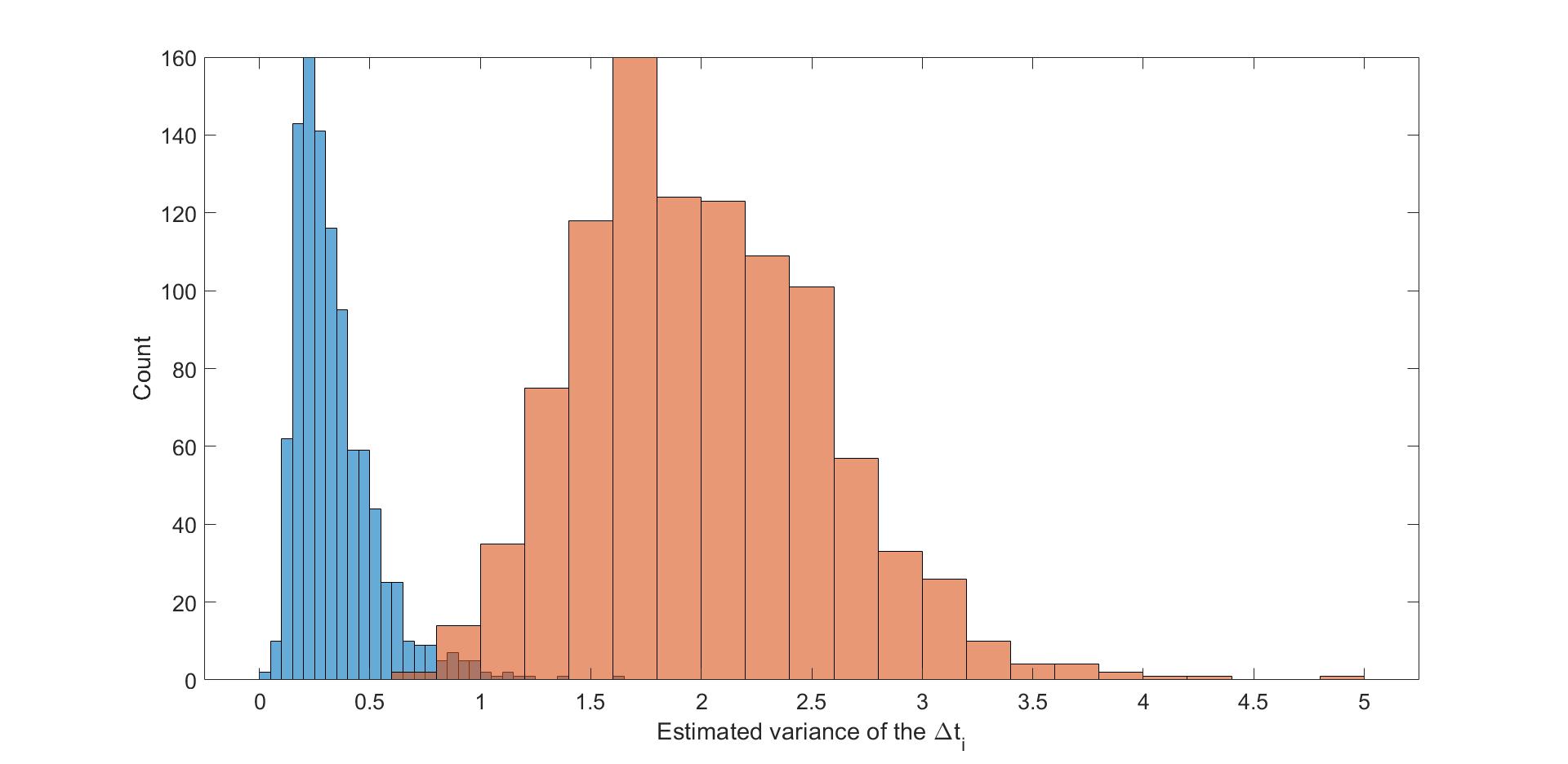}
    \caption{Empirical distribution of the estimated variance of the $\Delta t_i$ under simulated true detections (blue histogram), and under simulated false detections (red histogram).}
    \label{fig4}
\end{figure}

Following the simulation setup described above, 1000 true and 1000 false detections are simulated. For each detection, $\hat{\sigma_\epsilon}^2$ is then estimated using (\ref{var}). Figure \ref{fig4} shows the empirical distributions of $\hat{\sigma_\epsilon}^2$ for both true and false detections. The overlapping between distributions suggest that classification errors are possible.

\begin{figure}[ht!]
    \centering
     \captionsetup{}
    \includegraphics[width=1\textwidth]{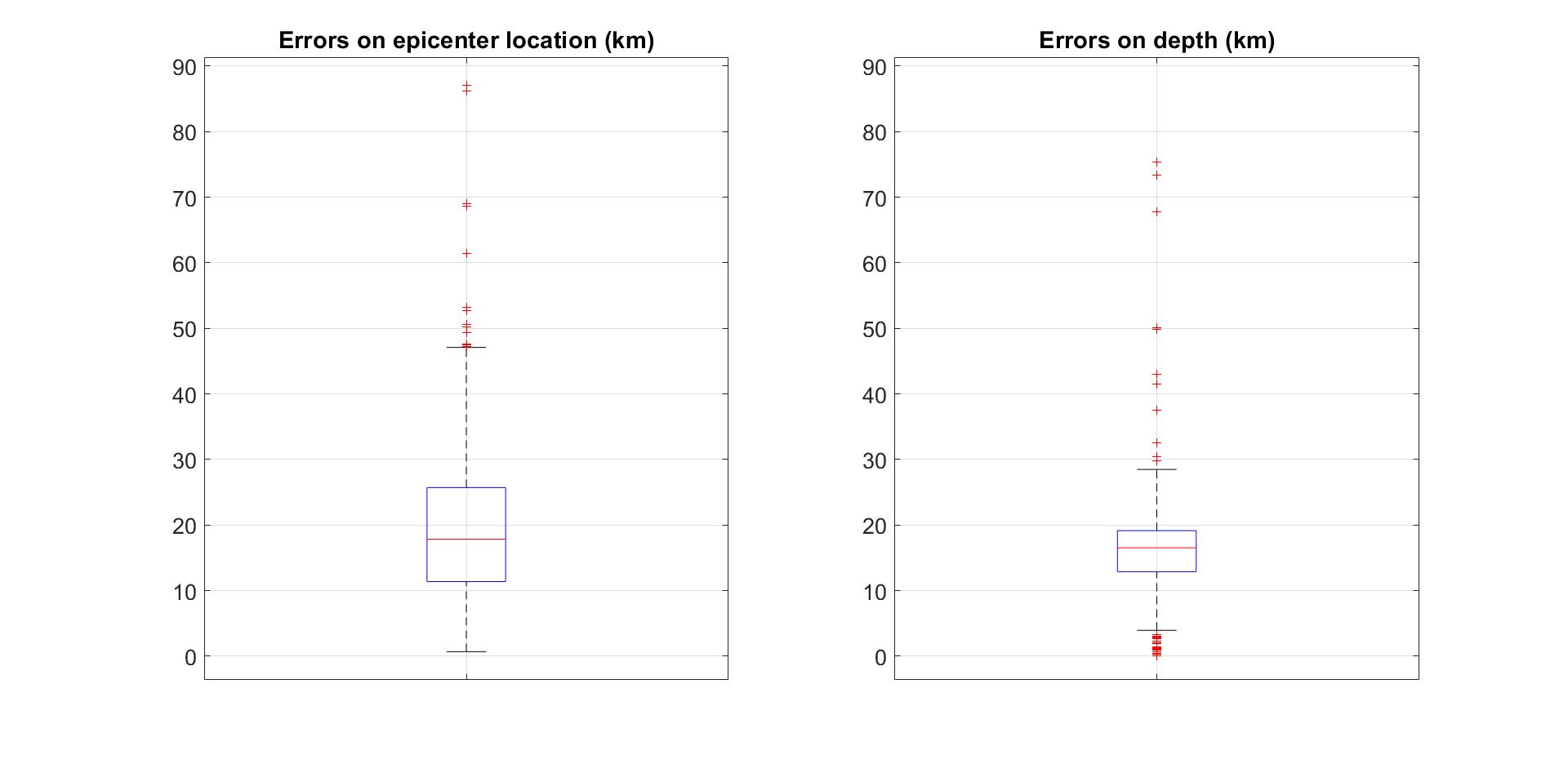}
    \caption{Box plot of the errors on epicentre location (left) and box plot of the errors on earthquake depth (right).}
    \label{fig5}
\end{figure}

Considering the empirical distribution for true detections, we found that fixing $\delta = 0.6$ in (\ref{test}) gives a type I error equal to $1\%$. Type II error is assessed using the empirical distribution for false detections and, for the same value of $\delta$, it is equal to $0.8\%$.

A by-product of detection classification are the estimates on the earthquake parameters. Figure \ref{fig5} show box-plots of errors on earthquake epicentre and depth. Both errors have a median of around $18$ km, suggesting that, along with the detection classification (true/false), the model output can be exploited to provide preliminary estimates of the earthquake parameters.

\section{Real data example}
\begin{figure}[ht!]
\captionsetup{}
    \centering
    \includegraphics[width=1\textwidth]{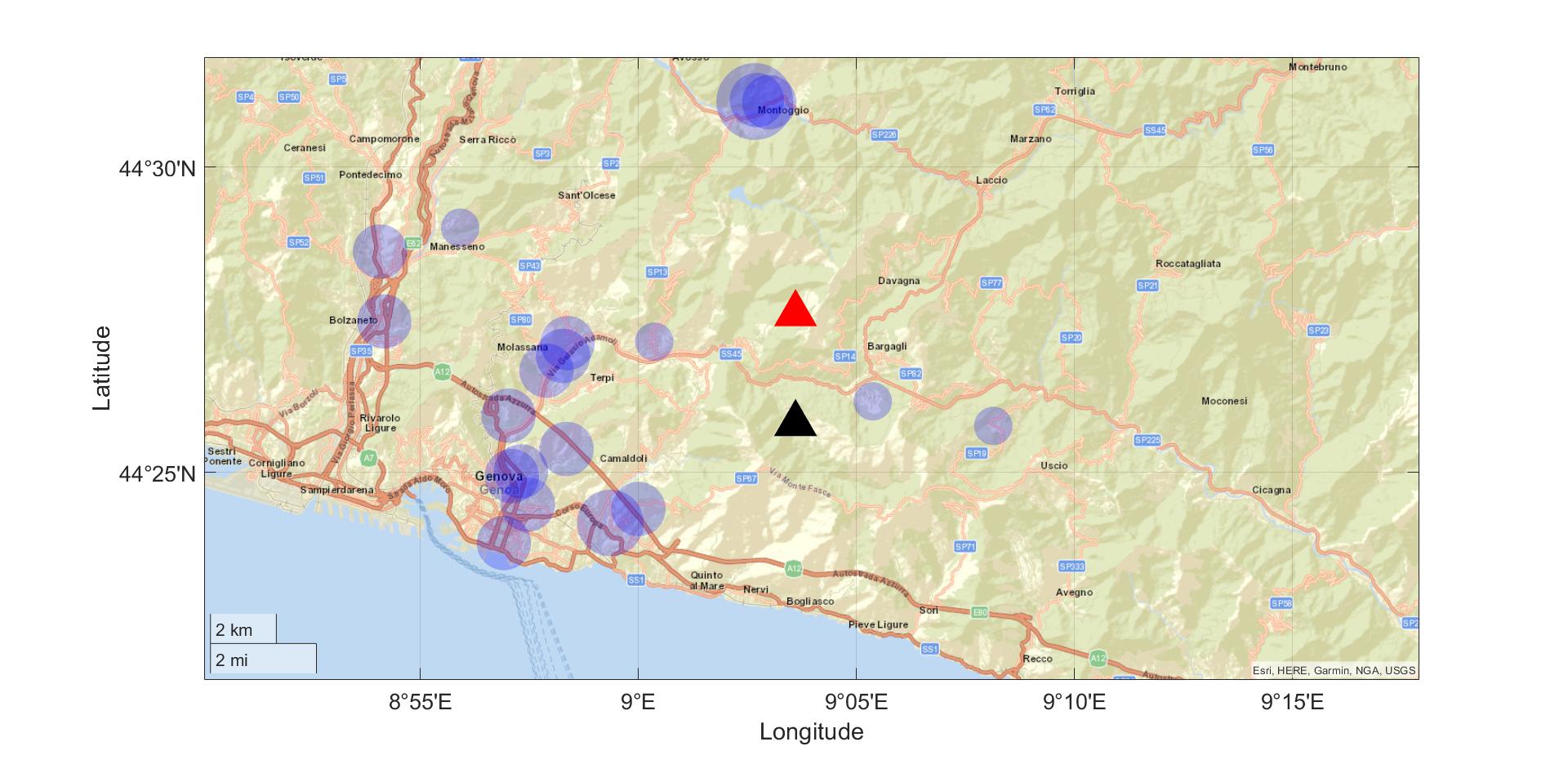}
    \caption{EQN triggers for the earthquake occurred on October 4, 2022 close to Genoa (Italy). Circles are triggering smartphone with radius proportional to the triggering time. The red triangle represent the true epicentre and the black triangle the estimated epicentre.}
    \label{fig:genova}
\end{figure}

The methodology developed in this work is applied on a true and on a false detections detection made by EQN. As a true earthquake, the event occurred near Genova (Italy) on October 4, 2022, at 21:41:10.5 UTC is considered. Figure \ref{fig:genova} depicts the triggering smartphones ($n=21$), while estimation and classification results are reported in Table \ref{tab:genova} for $v=7.8$ and $v=4.5$ km/s, respectively.

For both seismic wave velocities, we can observe that latitude and longitude are accurately estimated, while the error on depth is not negligible. Nonetheless, the true values are within the 95\% confidence intervals computed from the Hessian matrix given by the algorithm used to minimize (\ref{mle}). Additionally, the earthquake is classified as true under both velocities since both observed test statistics are lower than the test critical value. This happens because triggers are close to the epicentre, and primary and secondary seismic waves are nearly concurrent.

\begin{table}[ht!]
 \centering
    \begin{tabular}{l|r|rr|rr} 

      & & \multicolumn{2}{c}{$v = 7.8$ km/s} & \multicolumn{2}{c}{$v = 4.5$ km/s} \\

      & \textbf{Real}     & \textbf{Estimated} & \textbf{Error}  & \textbf{Estimated} & \textbf{Error}  \\
      \hline
      \hline
    \textbf{Latitude} ($^\circ$) & 44.46 & 44.43  & 0.03 & 44.43  & 0.02   \\
                &  & \small[44.38, 44.47]  &  & \small[44.40, 44.45] &\\
                 &    &  & & &\\ 
      \hline 
    \textbf{Longitude} ($^\circ$) & 9.06 & 9.06   & 0.00 & 9.03  & 0.03 \\
                &   & \small[9.01, 9.11] & & \small[9.03, 9.09] & \\
                 &    &  & & &\\
      \hline
    \textbf{Depth} (km)  & 8.00 & 0.01  & 7.99 & 0.01 & 7.99  \\
                &  & \small[0.00, 8.36]  &  & \small[0.00,  3.80] & \\
                 &    &  & & &\\
     \hline          
    \textbf{Estimated variance}   &  - & 0.57   & -  & 1.03   & -  \\
                &    &  & &  &\\
    \hline
    \textbf{Test statistic value}  & -  & 17.18   & -  & 31.02   & -  \\
               &    &  & & &\\
    \hline
    \textbf{Critical value}  &  - & 34.80   & -  & 34.80   & -  \\
               &    &  & &  & \\
    \hline
    \textbf{Classification}   &  - & True earthquake   & -  & True earthquake  & -  \\&
               &    &  & &  \\
    \end{tabular}
    \caption{Detection classification and earthquake parameters estimation for the EQN detection near Genova (Italy) assuming $v=7.8$ and $v=4.5$ km/s. The number of triggering smartphones is $n = 21$. In brackets, 99\% confidence intervals. Real earthquake parameters are taken from the website of the European-Mediterranean Seismological Centre \url{www.emsc-csem.org}.}
    \label{tab:genova}
\end{table}

\begin{figure}[ht!]
\captionsetup{}
    \centering
    \includegraphics[width=1\textwidth]{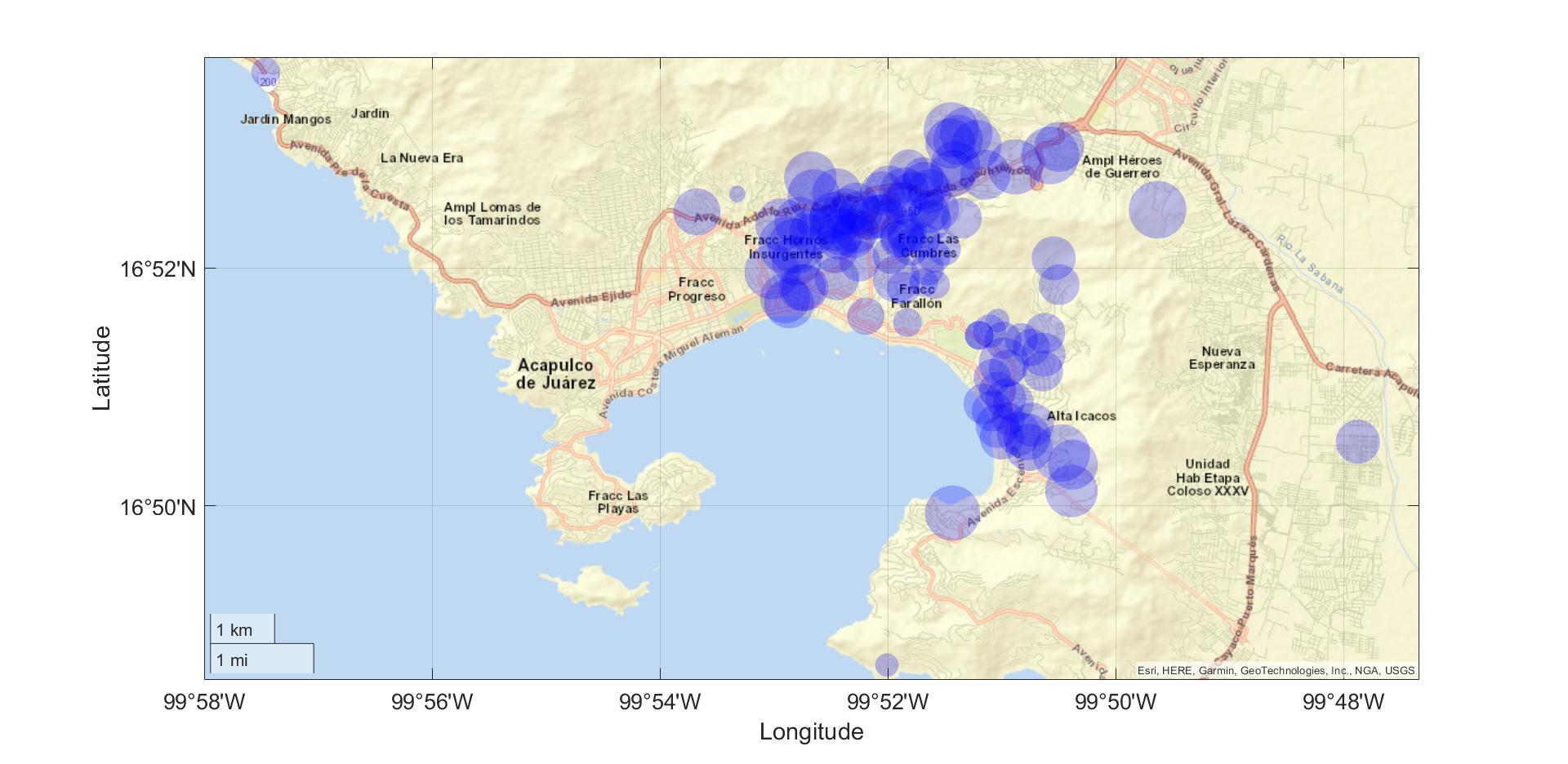}
    \caption{Triggers for the false EQN detection occurred on September 25, 2022, close to Acapulco (Mexico). Circles are triggering smartphone with radius proportional to the triggering time.}
    \label{fig:acapulco}
\end{figure}

The estimation and classification result was obtained in less than $1$ s using an Intel(R) Core(TM) i7-9750H CPU @2.60GHz, suggesting that the methodology is useful for real-time applications.

Figure \ref{fig:acapulco} shows, instead, the triggers of a false detection occurred near Acapulco (Mexico) on September 25, 2022, at 09:55:45 UTC. The detection is based on $n=108$ triggers. In this case, the computed test statistics are $1039.7$ and $1026.0$ for $v=4.5$ and $7.8$ km/s, respectively, while the critical values is $141.62$. $H_0$ is rejected in both cases and the detection is claimed as false.

\section{Conclusion}

This paper developed a statistical methodology for classifying earthquakes detected in real-time by smartphone-based earthquake early warning systems. The methodology is based on maximum likelihood estimation and on hypothesis testing. Thanks to its simplicity, classification and earthquake parameters estimation are performed in near real-time, making the methodology suitable to be implemented in operational systems. On the other hand, the methodology does not fully exploit the information content of the available data. In particular, the modelling is only on the triggering smartphones, while the active non-triggering smartphones are ignored. Knowing, at the earthquake detection time, which smartphones have not (yet) triggered may better constraint epicentre and depth, thus improving their estimates. This will be the focus of future works. 

\bibliographystyle{unsrt}
\bibliography{main}
\end{document}